\documentclass[a4paper,12pt]{article}
\pdfoutput=1
\usepackage{jcappub}
\usepackage{amsmath,amssymb}
\usepackage{subfigure}
\usepackage{color}
\usepackage{graphicx}
\usepackage{hyperref}
\usepackage{epstopdf}

\title{Helical phase inflation and its observational constraints}

\author[a]{Mudassar Sabir,}
\author[b]{Waqas Ahmed,}
\author[a,1]{Yungui Gong\note{Corresponding author},}
\author[c,d]{Tianjun Li,}
\author[a]{and Jiong Lin}

\affiliation[a]{School of Physics, Huazhong University of Science and Technology, Wuhan, Hubei 430074, China}
\affiliation[b]{School of Physics, Nankai University, Nankai District, Tianjin, China}
\affiliation[c]{School of Physical Sciences, University of Chinese Academy of Sciences, Beijing, China}
\affiliation[d]{CAS Key Laboratory of Theoretical Physics, Institute of Theoretical Physics, Chinese Academy of Sciences, Beijing 100190, China}

\emailAdd{msabir@hust.edu.cn}
\emailAdd{waqasmit@nankai.edu.cn}
\emailAdd{yggong@mail.hust.edu.cn}
\emailAdd{tli@itp.ac.cn}
\emailAdd{jionglin@hust.edu.cn}

\abstract{
We consider a class of helical phase inflation models from the ${\mathcal N}=1$ supergravity where the phase component of a complex field acts as an inflaton.
This class of models avoids the eta problem in supergravity inflation due to the phase monodromy of the superpotential.
We study the inflationary predictions of this class of models in the context of both standard and large extra dimensional brane cosmology, and find that they can easily accommodate the Planck 2018 and BICEP2 constraints.
We find that the helical phase inflation has $\alpha$-attractors
and the attractors depend on one model parameter only.
}

\begin{document}
\maketitle
\flushbottom

\section{Introduction}\label{sec:Intro}
Our observable universe has a finite age of 14 billion years, while it has already expanded to
about 46 billion light years. And it appears spectacularly homogeneous and isotropic with
cosmic microwave background temperature anisotropies only at the order of $10^{-5}$ or less.
The current far off regions of the universe were in causal contact, and the universe may have gone through
an era of exponential expansion which provides a solution to these puzzles
in standard cosmology~\cite{Guth:1980zm,Starobinsky:1980te,Sato:1980yn,Linde:1981mu,Albrecht:1982wi,Guth:1982ec}.
The inflationary models have specific predictions that can be tested experimentally.
A plethora of inflation models are available in the literature, see Ref.~\cite{Martin:2013tda} for a detailed list.
For slow-roll inflation, field excursions are related to the primordial gravitational wave
that has not been detected so far. However, the future satellite experiments
 will have the required sensitivity to measure the tensor-to-scalar ratio up to $\sim 0.001$.
 For example,
the future LiteBIRD, an experiment designed for the detection of B-mode polarization pattern embedded
in the Cosmic Microwave Background anisotropies, is sensitive enough to detect primordial gravitational waves up to $r\sim 10^{-3}$ \cite{Suzuki:2018cuy}.

Helical phase inflation from ${\mathcal N}=1$ supergravity was proposed where the phase of
a complex field acts as an inflaton while the radial component is strongly stabilized~\cite{Li:2015taa,Li:2014vpa,Li:2014unh}.
The phase field rolls down along the deformed helicoid shaped potential.
It is a generically difficult problem to generate a sufficiently flat scalar potential
in supergravity due to the exponential K\"ahler potential factor in the scalar potential.
To circumvent this eta problem usually additional symmetries are imposed. In helical phase inflation,
 we have the $U(1)$ phase monodromy of superpotential to circumvent this eta problem automatically.
Moreover, the helical model can easily interpolate between the natural inflation \cite{Freese:1990rb}
and the Starobinsky-like inflation \cite{Starobinsky:1980te} in a single potential.

The four-dimensional single field chaotic inflationary models are
in tension with the distance conjecture $\Delta\theta<1$ \cite{Ooguri:2006in}
because of the large value of the tensor-to-scalar ratio.
In the context of distance conjecture~\cite{Ooguri:2006in,Ooguri:2018wrx,Palti:2019pca},
although geodesic trajectories are quite well understood,
the generalization to non-geodesic trajectories is still not clear at a quantitative level \cite{Landete:2018kqf,Achucarro:2018vey}.
In view of the quantum gravity constraint, it seems a good idea to explore the non-geodesic trajectories
as they might be less constrained than geodesic ones,
but it is still an open question whether
the large field excursions for monodromic axions will
eventually be compatible with quantum gravity.

In braneworld scenario the tension between the distance conjecture
and the large value of tensor-to-scalar ratio is significantly
reduced due to the modified Friedmann equation
with a $\rho^2$ correction \cite{Jaman:2018ucm,Lin:2018kjm,Brahma:2018hrd,Safsafi:2018cua,Es-sobbahi:2018yfh,Bhattacharya:2019ryo,Jawad:2019hzo,Sabir:2019wel,Sabir:2019bsh}. This makes brane inflation in large extra dimensional scenario
an interesting possibility to explore further.
Braneworld scenario is also a prediction of brane gas cosmology \cite{Alexander:2000xv}
where our 4-dimensional spacetime is embedded in a higher dimensional bulk.
Therefore, it is interesting to study the helical phase inflation in braneworld scenario.

In this paper, we discuss the helical phase inflation in the braneworld scenario. In section \ref{sec:helical}, we briefly review the helical phase inflation from the ${\mathcal N}=1$ supergravity.
By varying the parameters, we can interpolate from natural inflation  to
Starobinsky-like inflation. We study the observational constraints on the model parameters,
and present the viable parameter space which is consistent with the Planck 2018
data and BICEP2 results \cite{Akrami:2018odb,Ade:2018gkx}.
It is found that the natural inflation
is marginally consistent with the observations at the $2\sigma$ level.
Next, we consider the helical phase inflation in the setup
of the large extra dimensional scenario where our four-dimensional
world is embedded in a five-dimensional space-time~\cite{Langlois:2000ns}.
we discuss the modified Friedmann equation, and then
study the observational constraints.
Similarly, we present the viable parameter space which is consistent with the Planck 2018
data and BICEP2 results as well.
In particular, the natural inflation is excluded by the observations.
While the Starobinsky-like inflation
provides the favorable central value for spectral index $n_s$, and the range of values of
tensor-to-scalar ratio $r$ spans the full experimental and theoretical estimate. The constraints on the temperature of reheating are also discussed.

\section{Helical phase inflation}
\label{sec:helical}

In helical phase inflation, the inflaton $\theta$, {\it i.e.}, the phase component of a complex field,
is a pseudo Nambu-Goldstone boson (PNGB)~\cite{Li:2015taa,Li:2014vpa,Li:2014unh}.
The potential of a complex field admits helicoid structure and the inflaton evolves along a local valley,
tracing a helical trajectory. The K\"{a}hler potential $K$ and the holomorphic superpotential $W$ are
\begin{eqnarray}
K &=& \Phi\bar{\Phi}+X\bar{X}-g(X\bar{X})^2,\\
W &=& a\frac{X}{\Phi}(\Phi^{\chi}-1), \qquad \chi = b+ i c.
\label{Weq}
\end{eqnarray}
It was thoroughly shown in Ref.~\cite{Li:2015taa} that the exponent $\chi$ has a geometrical origin
 associated with non-geometric flux compactification of Type IIB string theory.

The eta problem of supergravity theory is solved due to a global $U(1)$ symmetry of the K\"{a}hler potential
that introduces phase monodromy in the superpotential.
\begin{eqnarray}
\Phi\rightarrow e^{2\pi i}\Phi, \qquad  K\rightarrow K,  \qquad   W\rightarrow W+a\frac{X}{\Phi}\Phi^{\chi}(e^{2\pi i \chi}-1). \label{mon1}
\end{eqnarray}
The scalar potential is determined by the K\"{a}hler potential $K$ and superpotential $W$
\begin{eqnarray}
V=e^K(K^{i\bar{j}}D_iW D_{\bar{j}}\bar{W}-3W\bar{W}),
\end{eqnarray}
where $K_{i\bar{j}}=\partial_i\partial_{\bar{j}}K$ and $D_iW=\partial_iW+K_iW$. During inflation the field $X$ is strongly fixed at its vacuum expectation value $\langle X\rangle=0$. Hence the scalar potential can be written as follows
\begin{eqnarray}
V(r,\theta)=a^2\frac{e^{r^2}}{r^2}\left(1+r^{2b}e^{-2c\theta}-2 r^b e^{-c\theta}\cos(b\theta+c\log r)\right).
\label{eq:V}
\end{eqnarray}
in reduced Planck units ($M_P\equiv 1/(8\pi G)=1$), with $\Phi\equiv re^{i\theta}$.

The norm of $\Phi$ needs to be stabilized otherwise it will generate notable iso-curvature perturbation that contradicts with observations.
However, it is a non trivial task to stabilize the norm of $\Phi$ while keep the phase light as the norm and phase couple to each other.
Curvature along the radial direction is determined by the coefficient $e^{r^2}/r^2$, which admits a global minimum at $r = 1$ and gives a large mass above the
Hubble scale. Extra couplings between the field norm $r$ and the phase $\theta$ in $V (r, \theta)$ can partially affect the stabilization of $|\Phi|$,
although $b \ll 1$ and $e^{-c\,\theta} \ll 1$ during inflation, such corrections to observables are of order ${\cal O}(b^2)$ and can be ignored.
With stabilized field norm $|\Phi|=1$ and $b\ll 1$, but no constraint on $c$, the scalar potential $V(r,\theta)$ becomes
\begin{equation}
\label{eq:V1}
V(\theta)= a^2\left[1+e^{-2c\theta}-2e^{-c\theta}\cos(b\theta)\right].
\end{equation}
By varying $b$ and $c$ we can interpolate from natural inflation ($c=0,\ b\ll1$) to Starobinsky-like inflation ($b=0,\, c>0$).

Let us compare our model with the usual $\alpha$-attractor models. From no-scale like K\"ahler potential,
one obtains two kinds of $\alpha$-attractor models: the T-Model and E-Model \cite{Kallosh:2015lwa}.
The potential for the T-Model is
\begin{equation}
\label{T-Models}
V(\phi) = \alpha \mu^2 \tanh^2\frac{\phi}{\sqrt {6\alpha}}.
\end{equation}
In general, our model is different from
the usual $\alpha$-attractor T-models derived from
no-scale like K\"ahler potential~\cite{Kallosh:2015lwa}.
If $e^{\frac{2\phi}{\sqrt {6\alpha}}}$ is much larger than 1,
we can obtain the T-model from our model \eqref{eq:V1} by choosing
\begin{equation}
\label{Conditions1}
a^2= \alpha \mu^2~,~~b=0~,~~c=\sqrt{\frac{2}{3\alpha}}~,~ \theta=\phi-\ln2/c.
\end{equation}
The inflaton potential for the E-Model is
\begin{equation}
\label{E-Models}
V(\phi) = \alpha \mu^2 \left(1-e^{-\sqrt{\frac{2}{3\alpha}} \phi}\right)^2,
\end{equation}
and we can obtain the above inflaton potential for E-Model from our model \eqref{eq:V1} by choosing
\begin{equation}
\label{Conditions2}
a^2= \alpha \mu^2~,~~b=0~,~~c=\sqrt{\frac{2}{3\alpha}}~,~~\theta=\phi.
\end{equation}
Therefore, our model is more general than
the usual $\alpha$-attractor T-models
and E-models from no-scale like K\"ahler potential \cite{Kallosh:2015lwa}.

\begin{figure}[ht]
\centering
\subfigure{\includegraphics[width=0.45\textwidth]{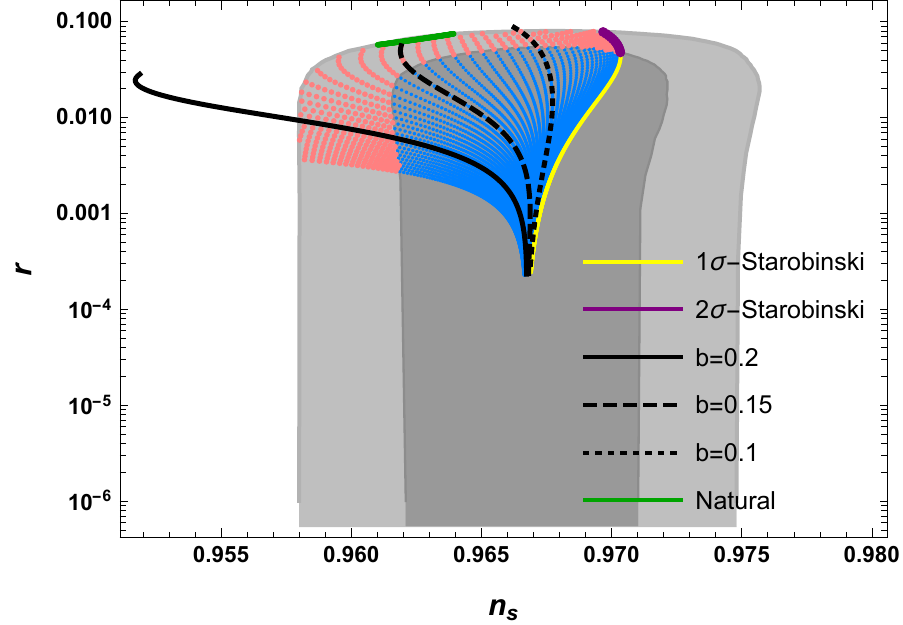}}\qquad
\subfigure{\includegraphics[width=0.45\textwidth]{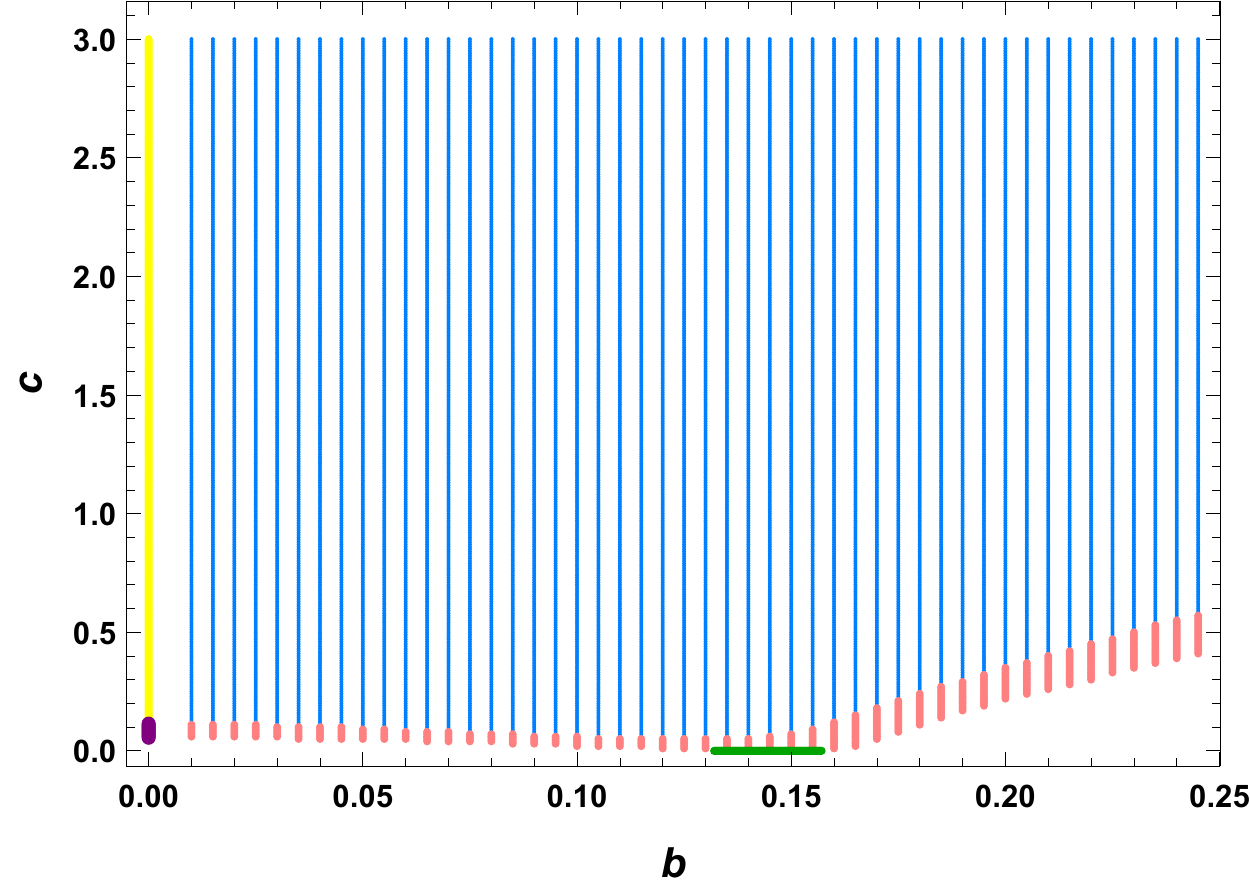}}
\caption{The constraints on helical phase inflation with $N=60$
$e$-folds in GR case. The light blue dots correspond to the $1\sigma$ constraints and
the pink dots correspond to the $2\sigma$ constraints.
The shaded regions are the marginalized $1\sigma$ and $2\sigma$ contours from Planck 2018 and BICEP2 results \cite{Akrami:2018odb,Ade:2018gkx}. In the left panel, the $\alpha$-attractor behaviours are shown for several different values of $b$, the value of $r$ decreases as $c$ increase. The right panel shows the $1\sigma$ and $2\sigma$ constrains on the model parameters $b$ and $c$.}
\label{Fig:1}
\end{figure}

In Fig. \ref{Fig:1}, we show the numerical results of the scalar spectral index $n_s$ and the tensor-to-scalar ratio $r$ for the helical phase inflation in the framework of general relativity (GR).
By comparing with the Planck 2018 and BICEP2 observations \cite{Akrami:2018odb}, we get the constraints on the parameters of the model.
Note that $b$ and $c$ control the period and amplitude of potential, respectively.
As $b$ increases, the potential becomes steep and large $c$ is needed to flatten the potential and realize inflation.
In the limit of $b\theta \ll 1$ and $c\theta \ll 1$, we get the polynomial potential
\begin{equation}
V\simeq \frac{1}{2}a^2b^2\theta^2(1-c\theta).
\end{equation}
This potential is excluded by Planck 2018 and BICEP2 observations as shown in Fig. \ref{Fig:1}.

For the Starobinsky-like inflation ($b=0$) the number of $e$-folds,
scalar spectral index and tensor-to-scalar ratio are
\begin{eqnarray}
N_* &=& \frac{e^{c\,\theta_*}-e^{c\,\theta_e}-c(\theta_*-\theta_e)}{2c^2},\\
n_s &= & 1-\frac{4 c^2 \left(e^{c\,\theta_* }+1\right)}{\left(e^{c\,\theta_* }-1\right)^2},\\
r &=& \frac{32\,c^2}{(e^{c\, \theta_{*}}-1)^2},
\end{eqnarray}
where $\theta_e$ is the field value evaluated when inflation ends
and can be solved from $\epsilon(\theta_e)=1$,
\begin{eqnarray}
c\,\theta_e = \ln(1+\sqrt{2}c).
\end{eqnarray}
In the limit $c\gg 1$, $N_*\simeq(e^{c\,\theta_*}-e^{c\,\theta_e})/(2c^2)$,
we get the $\alpha$-attractor result \cite{Kallosh:2013yoa}
\begin{equation}
\label{eq:ns_GR}
\begin{split}
n_s &\simeq 1-\frac{4c^2}{e^{c\,\theta_*}}\,\,\simeq\,\, 1-\frac{2}{N_*},\\
r &\simeq \frac{32c^2}{e^{2c\,\theta_*}}\,\,\simeq\,\, \frac{8}{c^2N_*^2}.
\end{split}
\end{equation}
As discussed in Eqs. \eqref{Conditions1} and \eqref{Conditions2},
we can obtain T-models and E-models if we identify $c=\sqrt{2/(3\alpha)}$,
thus it is no surprise that we obtain the $\alpha$-attractors
and large $c$ plays the role of suppressing the tensor-to-scalar ratio.
The results are shown in Fig. \ref{Fig:1} with yellow and green lines.

In the limit $c\theta\gg 1$ with $b\neq 0$, the potential becomes
\begin{equation}
  V\simeq a^2\left( 1-2\mathrm{cos}(b\theta)e^{-c\theta}\right),
\end{equation}
Note that as $c$ gets larger, the field value decreases and $b\theta$ becomes small. Thus when $c\gg 1$, the number of $e$-folds before the end of inflation can be expanded in terms of $b\theta$ as
\begin{equation}
  N_*\simeq-\int_{\theta_*}^{\theta_{\rm e}}\frac{e^{c\theta}}{2c\mathrm{cos}(b\theta)}d\theta
  \simeq -\int_{\theta_*}^{\theta_{\rm e}}\frac{e^{c\theta}}{2c}
  (1+b^2\theta^2/2)d\theta
  \simeq \frac{e^{c\theta_*}(1+b^2\theta_*^2/2)}{2c^2},
\end{equation}
the scalar spectral index and tensor-to-scalar ratio are
\begin{equation}
\begin{split}
n_s&\simeq 1-\frac{4c^2\mathrm{cos}(b\theta_*-\theta_0)}{e^{c\theta_*}}
  \simeq 1-\frac{4c^2}{e^{c\theta_*}}(1-b^2\theta_*^2/2)
  \simeq 1-\frac{2}{N_*}, \\
r &\simeq  \frac{32c^2\mathrm{cos}^2(b\theta_*-\theta_1)}{e^{2c\theta_*}}
\simeq \frac{32c^2}{e^{2c\theta_*}}(1-b^2\theta_*^2/2)^2
\simeq\frac{8}{c^2N_*^2},
\end{split}
\end{equation}
where $\theta_0=\mathrm{arctan}(2b/c)\approx 0$ and $\theta_1=\mathrm{arctan}(b/c)\approx 0$.
The attractors are the same as $\alpha$-attractors \eqref{eq:ns_GR} in Starobinsky-like case
and are independent of the value of $b$ in the large $c$ limit if we identify $c=\sqrt{2/(3\alpha)}$.
The $\alpha$-attractors are shown in Fig. \ref{Fig:1}.

For natural inflation ($c=0$), the corresponding analytic expressions
for the number of $e$-folds, spectral index and tensor-to-scalar ratio are
\begin{eqnarray}
N_* & = &\frac{2}{ b^2} \ln \left[\frac{\cos \left(b\,\theta_{e} / 2 \right)}{\cos \left(b\,\theta_{*} / 2\right)}\right],\\
n_s &= & -2 b^2 \csc ^2\left(\frac{b\,\theta_{*} }{2}\right)+b^2+1,\\
r &=& 8 b^2 \cot ^2\left(\frac{b\,\theta_{*} }{2}\right),
\end{eqnarray}
and the end of inflation condition $\epsilon(\theta_e)=1$ determines the field $\theta_e$ as
\begin{eqnarray}
b\,\theta_e=\arccos\left(\frac{2-b^2}{2+b^2}\right).
\end{eqnarray}
The results are marginally consistent with the observations
at the $2\sigma$ confidence level as shown with cyan line in Fig. \ref{Fig:1}.
The field excursions in four-dimensional case are typically of the order of $\sim {\cal O}(10)$
in reduced Planck units as elaborated in Ref.~\cite{Li:2015taa}.

\section{Brane inflation}
\label{sec:brane}

In the braneworld cosmology, our four-dimensional world is a 3-brane embedded in a higher-dimensional bulk.
The Friedmann equation is modified due to high energy corrections to
Einstein equations on the brane \cite{Shiromizu:1999wj,Csaki:1999jh,Maartens:1999hf,Binetruy:1999hy,Binetruy:1999ut}
\begin{equation}
H^2=\frac{\rho}{3M_P^2}\left(1+\frac{\rho}{2 \lambda} \right),
\label{eq:Friedmann}
\end{equation}
where $\rho$ is the energy density of the scalar field, $M_P=M_4/\sqrt{8\pi}$ is the reduced Planck mass,
and $\lambda$ is the brane tension that relates the four-dimensional Planck scale $M_4$
with the five-dimensional Planck scale $M_5$ as below
\begin{equation}
\lambda=\frac{3}{4\pi} \frac{M_5^6}{M_4^2}.
\end{equation}
The nucleosynthesis limit implies that $\lambda \gtrsim (1 \mbox{ MeV})^4 \sim (10^{-21})^4$ in the reduced Planck unit.
A more stringent constraint can be obtained by requiring the theory to be reduced to Newtonian gravity on scales
larger than 1 mm corresponding to $\lambda \gtrsim 5 \times 10^{-53}$, i.e.,
$M_5 \gtrsim 10^5$ TeV~\cite{Maartens:1999hf}. Notice that in the limit $\lambda \rightarrow\infty$,
we recover the standard Friedman equation in four dimensions.

The $\rho^2$ correction term in the modified Friedmann equation
makes the slow-roll parameters smaller for a given potential such that~\cite{Maartens:1999hf,Langlois:2000ns}
\begin{equation}
\begin{split}
\epsilon_H&=\epsilon_V\frac{1+V/\lambda}{\left[1+V/(2\lambda)\right]^2},\\
\eta_H&=\eta_V\frac{1}{1+V/(2\lambda)},
\end{split}
\end{equation}
where $\epsilon_V=(V'/V)^2/2$ and $\eta_V=V''/V$. Slow-roll brane inflation can be realized when $\epsilon_H\ll1$ and $\eta_H\ll1$.
The number of $e$-folds during inflation becomes
\begin{equation}
N_* = -  \int_{\theta_*}^{\theta_{\rm e}}\frac{ V }{ V'} \left(
1+\frac{ V}{ 2\,\lambda}\right) d{\theta}.
\label{eq:efolds_B}
\end{equation}
Note that in the high energy limit $V\gg\lambda$, to get the
same number of $e$-folds, brane inflation
requires smaller field excursion than that in GR.
For the Randall-Sundrum model II \cite{Randall:1999vf},
the amplitudes for the tensor and scalar power spectrum are \cite{Maartens:1999hf, Langlois:2000ns}
\begin{eqnarray}
A_t^2 &=&\frac{2}{3 \,\pi^2}\,  V \, \left( 1+\frac {V}{ 2\, \lambda}\right)F^2, \label{eq:P_t}\\
A_s^2 &=& \frac{1}{12 \,\pi^2}\,\frac{V^3}{V^{\prime2}} \left( 1 + \frac{V}{2\,\lambda} \right)^3 ,
\label{eq:P_s}
\end{eqnarray}
where
\begin{equation}
F^2=\left[\sqrt{1+x^2} -x^2 \sinh^{-1}\left(\frac{1}{x}\right)\right]^{-1},
\label{eq:F2}
\end{equation}
with
\begin{align}
\label{eq:xV}
x\equiv \left(\frac{3 \,H^2}{4 \pi\,\lambda}\right)^{1/2} = \left[\frac{2\,V} {\lambda}\left(1+\frac{V}{2\,\lambda}\right)\right]^{1/2}.
\end{align}
 Note that the right-handed sides of
Eqs. (\ref{eq:P_t}) and (\ref{eq:P_s}) should be evaluated at the horizon crossing.
In the low-energy limit $V/\lambda\ll 1$, $F^2\approx 1$, we recover
the results in standard cosmology\footnote{The results \eqref{eq:P_t} and \eqref{eq:P_s} differ from those in Refs. \cite{Maartens:1999hf,Langlois:2000ns}
by factors of $(1/16)\times (4/25)$ and $4/25$ respectively because they use matter density perturbations when modes re-enter the Hubble scale during the matter dominated era.}.

The scalar spectral tilt and the tensor-to-scalar ratio under slow-roll approximation can be written as \cite{Maartens:1999hf} \cite{Bento:2008yx}
\begin{equation}
\begin{split}
   &n_{s} - 1 \simeq -6\epsilon_H + 2\eta_H, \\
   &r = \frac{A_t^2}{A_s^2}
\simeq8\left(\frac{V'}{V}\right)^2\frac{F^2}{[1+V/(2\lambda)]^2}.
\end{split}
\end{equation}
In the low energy limit $V/\lambda\ll 1$, $F^2\approx 1$, we recover the standard result $r\simeq16\epsilon_V$. In the high energy limit where $V/\lambda\gg 1$ and $F^2\approx 3\,V/2\,\lambda$,
the tensor-to-scalar ratio and the scalar power spectrum amplitude become
\begin{equation}
\label{braseq}
\begin{split}
r&\simeq24\epsilon_H,\\
A_s^2&\simeq \frac{1}{12 \,\pi^2}\,\frac{V}{\epsilon_H} \left( \frac{V}{2\,\lambda} \right)^2.
\end{split}
\end{equation}
Hence in brane inflation, the inflation energy scale and the brane tension cannot be completely  fixed by the observational values of $A_s^2$ and $r$.

For the Starobinsky-like case with $b=0$, the analytic expressions for $N_*$, $n_s$ and $r$
in the high energy and large $c$ limits are
\begin{equation}
\label{brne}
N_* \simeq \frac{a^2}{4\,\lambda \,c^2} e^{c\,\theta_*},
\end{equation}
\begin{equation}
\label{brnseq1}
\begin{split}
n_s &\simeq 1-\frac{8\,\lambda \,c^2}{a^2}\frac{1}{e^{c\, \theta_*}}\,\,\simeq\,\, 1-\frac{2}{N_*},\\
r &\simeq \frac{192\,\lambda \,c^2}{a^2}\frac{1}{e^{2c\, \theta_*}} \,\,\simeq\,\, \frac{12}{c^2N_*^2}.
\end{split}
\end{equation}
The attractors \eqref{brnseq1} are the same as
$\alpha$-attractors \eqref{eq:ns_GR} in GR
except that in brane case $r$ is $3/2$ times larger. Comparing with E-model $\alpha$-attractors in braneworld \cite{Sabir:2019wel} in the large $N$ limit, again we have the relation $c=\sqrt{2/(3\alpha)}$.

For the general case with $b\neq 0$,
in the large $c$ and high energy limits, $c\theta\gg1$ and $V\gg\lambda$, $b\theta$ is small and we can expand the observables in terms of $b\theta$.
The number of $e$-folds is
\begin{equation}
  N_*\simeq-\int_{\theta_*}^{\theta_{\rm e}}\frac{a^2e^{c\theta}}{4c\lambda\mathrm{cos}(b\theta)}d\theta
  \simeq -\frac{a^2}{2\lambda}\int_{\theta_*}^{\theta_{\rm e}}\frac{e^{c\theta}}{2c}
  (1+b^2\theta^2/2)d\theta
  \simeq\frac{a^2e^{c\theta_*}(1+b^2\theta_*^2/2)}{4\lambda c^2}.
\end{equation}
The scalar spectral index and tensor-to-scalar ratio are
\begin{equation}
\label{brnseq2}
  \begin{split}
     n_s & \simeq 1-\frac{8\lambda c^2\mathrm{cos}(b\theta_*)}{a^2e^{c\theta_*}}
     \simeq1-\frac{8\lambda c^2}{a^2e^{c\theta_*}}(1-b^2\theta_*^2/2)
     \simeq1-\frac{2}{N_*}, \\
      r & \simeq \frac{192\lambda c^2\mathrm{cos}^2(b\theta_*)}{a^2e^{2c\theta_*}}
      \simeq \frac{192\lambda c^2}{a^2e^{2c\theta_*}}(1-b^2\theta_*^2/2)^2
      \simeq\frac{12}{c^2N_*^2}.
  \end{split}
\end{equation}
Therefore, the attractors for $b\neq 0$ are the same as those with $b=0$ and are independent of the value of $b$ in the large $c$ limit.

\begin{figure}[htp]
\centering
\subfigure{\includegraphics[width=0.45\textwidth]{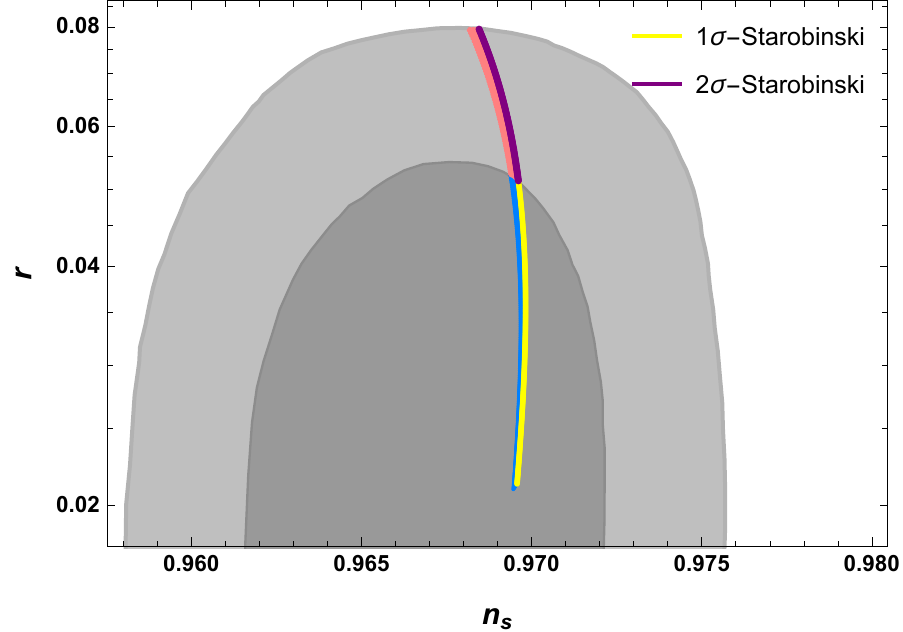}}\qquad
\subfigure{\includegraphics[width=0.45\textwidth]{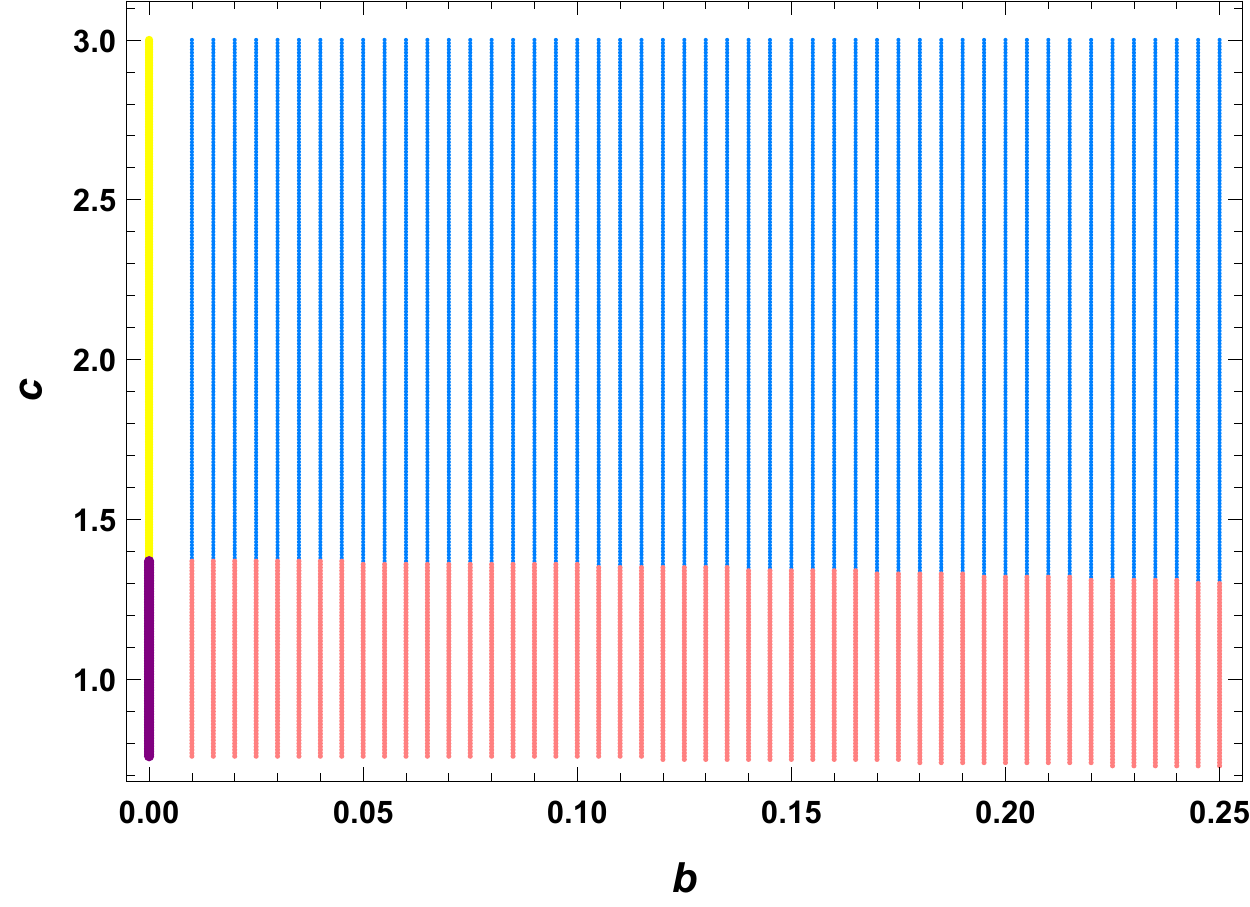}}
\caption{The observational constraints on helical phase inflation on a brane taking $N=60$ and $a^2/\lambda =100$. The left panel shows the attractors. The $1\sigma$ and $2\sigma$ constraints on the parameters $b$ and $c$ are shown with light blue and pink colors respectively in the right panel. }
\label{Fig:2}
\end{figure}

\begin{figure}[htp]
\centering
\includegraphics[width=0.6\textwidth]{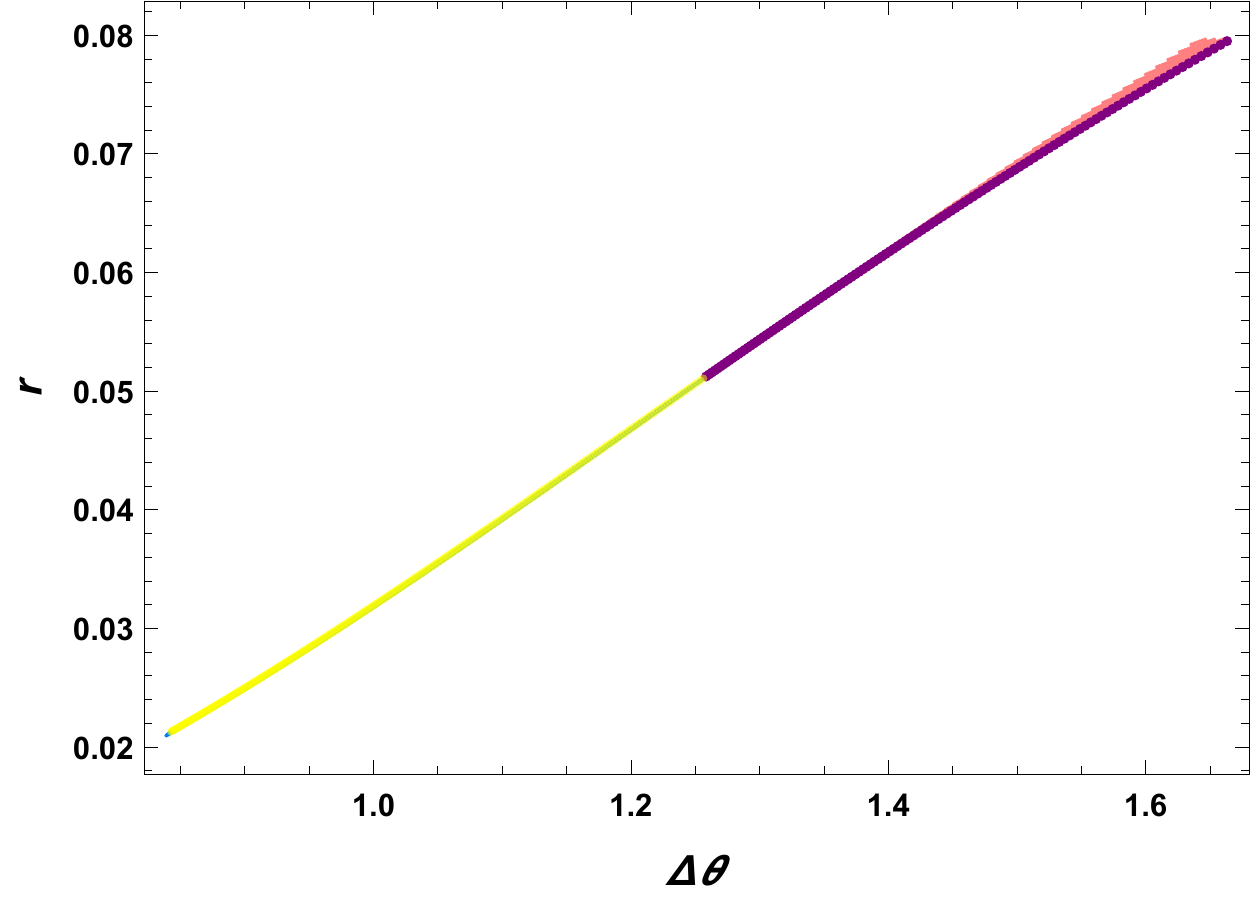}
\caption{The results for $r$ and $\Delta\theta$ by using the constrained parameters in Fig. \ref{Fig:2}.}
\label{Fig:3}
\end{figure}

Fig. \ref{Fig:2} shows the numerical results of the scalar spectral index $n_s$ and the tensor-to-scalar ratio $r$ in brane helical phase inflation for $N=60$ $e$-folds and the ratio of scale of inflation to the brane tension $a^2/\lambda=100$.
Because we are interested in the effect of brane correction,
so we choose the high energy limit $a^2/\lambda \gg 1$.
The predictions of the $n_s-r$ plane shrink from a tree-like
shape ($b\neq 0$) to a single branch ($b=0$) as we increase the value of $a^2/\lambda$,
here we take a relatively large value $a^2/\lambda=100$. A larger value of $a^2/\lambda$ just gives the Starobinski-like branch as derived in Eq. \eqref{brnseq2}.
By comparing with the Planck 2018 and BICEP2 observations, we get the constraints on the model parameters $b$ and $c$.
The above attractors \eqref{brnseq1} and \eqref{brnseq2} are confirmed from the numerical results in Fig. \ref{Fig:2}.
Note that due to the tensor-to-scalar ratio $r$ receives the correction $F^2/[1+V/(2\lambda)]^2$, in the high energy limit
the tensor-to-scalar ratio $r$ is larger in brane inflation than that in GR. Because of that, larger values of $c$ are needed.
The same reason leads to the exclusion of natural inflation on a brane by the observations.
Due to the high energy correction factor $V/\lambda$ in brane inflation,
as shown in Fig. \ref{Fig:2} attractors can be reached more easily and the observables $n_s$
and $r$ are almost independent of $b$.
By using constrained parameters $b$ and $c$ in Fig. \ref{Fig:2}, we also calculate
the field excursion $\Delta\theta$ and the results for $r$ and $\Delta\theta$ are shown in Fig. \ref{Fig:3}. Thus,
the sub-Planckian field excursion $\Delta\theta <1$ is obtained
if $r<0.03$.

\subsection{Reheating}
\label{sec:Reheating}

\begin{figure}[htp]
\centering
\includegraphics[width=0.6\textwidth]{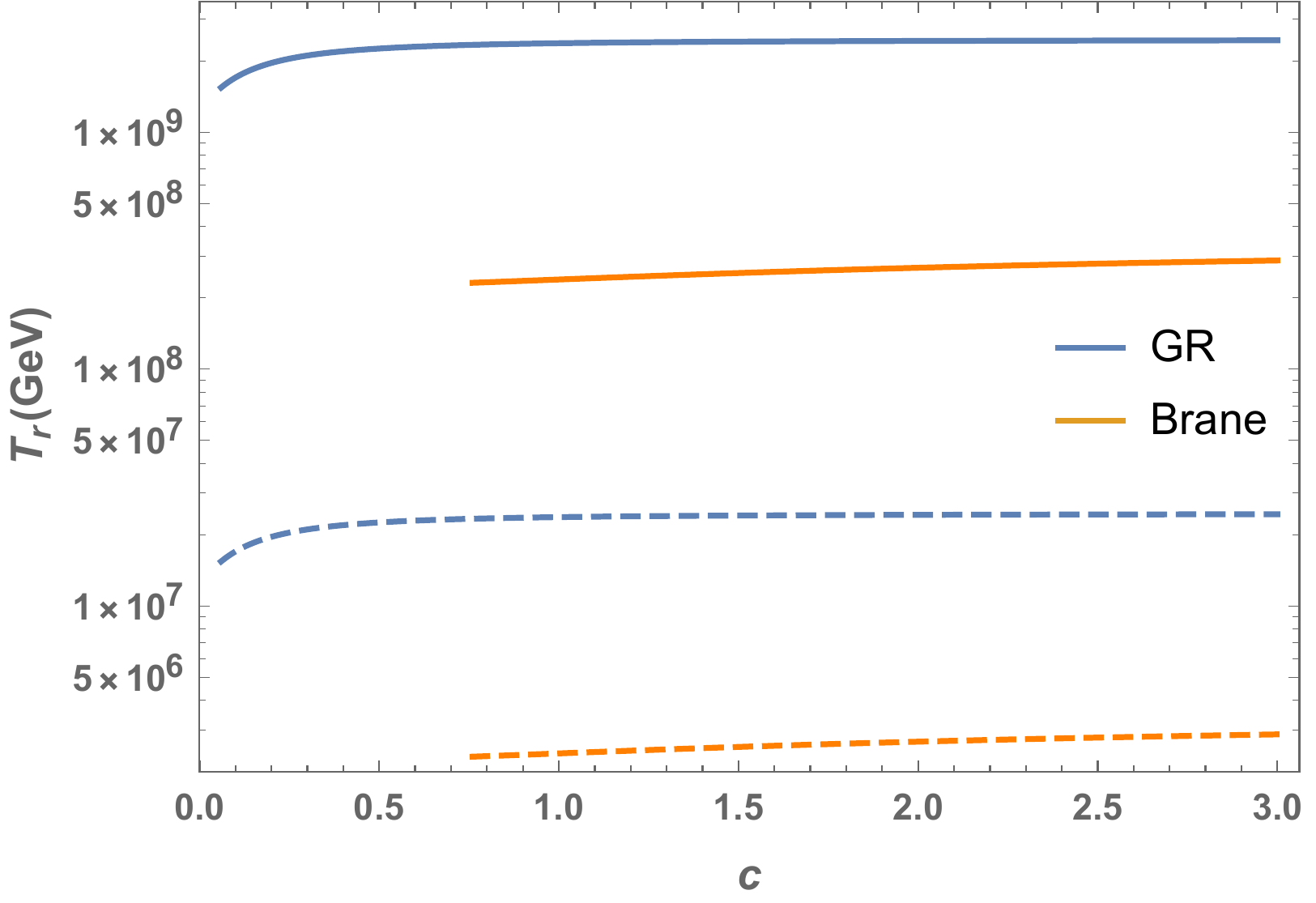}
\caption{Plots of reheating temperature versus parameters $c$ for Starobinsky-like helical phase inflation where we have set the ratio $a^2/\lambda =100$. The solid lines correspond to $\gamma=10^{-3}$ and the dashed lines correspond to $\gamma=10^{-5}$.}
\label{fig:4}
\end{figure}
In order to calculate the reheating temperature we add an interaction term in superpotential as,
\begin{eqnarray}
W & \supset & \gamma\, \Phi H_uH_d
\end{eqnarray}
that induces an inflaton decay into Higgsinos via the decay width given by \cite{Okada:2015vka},
\begin{equation}
 \Gamma_\Phi(\Phi \rightarrow \tilde H_u\tilde H_d)=\frac{\gamma ^2 }{8 \pi }m_{\Phi},
\end{equation}
where $\gamma$ is the decay coupling parameter and $m_\Phi=2 a^2 \left(b^2+c^2\right)$ is the inflaton mass. The reheating temperature $T_r$ is estimated to be \cite{Kolb:1990vq}
\begin{eqnarray}\label{reheat}
T_r & \approx &\sqrt[4]{\frac{90}{\pi^2 g_*}} \sqrt{\Gamma_\Phi }\,,\\
&=& 3.025\times 10^{17} \gamma  \sqrt{a^2 \left(b^2+c^2\right)}\,{\rm GeV}.
\end{eqnarray}
where $g_*$ is taken to be 228.75 for MSSM. Figure \ref{fig:4} shows plots of reheating temperature versus parameter $c$ for the case of Starobinsky-like inflation with $b=0$ and $N=60$ $e$-folds. The solid lines correspond to $\gamma = 10^{-3}$ and the reheating temperature is $T_r\simeq10^{9}$ GeV. The LHC bounds on gravitino mass constraints the reheating temperature to be $T_r\lesssim 10^{9}$ GeV \cite{Ahmed:2018jlv}.
The dashed lines correspond to $\gamma=10^{-5}$ and the
reheating temperature is $T_r\simeq10^{6}$ GeV. Due to the correction
factor $(1+V/2\lambda)^3$ to scalar perturbation amplitudes of brane inflation,
the reheating temperature for brane inflation is always less than GR inflation
in the high energy regime $V/\lambda\gg 1$.

\section{Conclusions}
\label{sec:Conclusions}
We have considered the helical phase inflation models from the ${\mathcal N}=1$ supergravity
where the phase component of a complex field is inflaton.
This class of models can solve the eta problem in supergravity inflation due to the phase monodromy of the superpotential.

We studied the observational constraints on the model parameters, and
 present the viable parameter space which is consistent with the Planck 2018 and BICEP2 results.
In GR, the natural inflation is marginally consistent with the observations at the $2\sigma$ level. We also find that the helical phase inflation has $\alpha$-attractors in the large $c$ limit
and the attractors are independent of the parameter $b$.
In the brane case,
the natural inflation lies well outside the 2$\sigma$ region in $n_s$-$r$ plane and it is excluded by the observations.
The $\alpha$-attractors are also found except that the value
of $r$ is $3/2$ times larger than that in GR. The high energy correction makes the attractors easily reached and the attractors
for the general case with $b\neq 0$ is almost the same as the
Starobinsky-like case with $b=0$. The field excursions in the brane scenario can be subplanckian if $r<0.03$.

From the reheating analysis in Sec. \ref{sec:Reheating}, the brane inflation having an additional parameter can easily accommodate a wide range for the reheating temperature. The upper bound for the temperature can  be lowered with either taking the coupling to Higgsinos to be small or taking a higher ratio of scale of inflation to the brane tension.

\acknowledgments

MS was supported by Higher Education Commission Pakistan Ph.D. scholarship.
This research was supported  by  the  National  Natural  Science  Foundation  of  China under Grant Nos. 11647601, 11875062, and
11875136,  by  the
Major Program of the National Natural Science Foundation of China under Grant
No.~11690021, and by the Key Research Program of Frontier Science, CAS.

%\bibliographystyle{JHEP}
%\bibliography{References}

\begin{thebibliography}{10}

\bibitem{Guth:1980zm}
A.~H. Guth, \emph{{The Inflationary Universe: A Possible Solution to the
  Horizon and Flatness Problems}},
  \href{https://doi.org/10.1103/PhysRevD.23.347}{\emph{Phys. Rev. D} {\bfseries
  23} (1981) 347}.

\bibitem{Starobinsky:1980te}
A.~A. Starobinsky, \emph{{A New Type of Isotropic Cosmological Models Without
  Singularity}},
  \href{https://doi.org/10.1016/0370-2693(80)90670-X}{\emph{Phys. Lett. B}
  {\bfseries 91} (1980) 99}.

\bibitem{Sato:1980yn}
K.~Sato, \emph{{First Order Phase Transition of a Vacuum and Expansion of the
  Universe}}, {\emph{Mon. Not. Roy. Astron. Soc.} {\bfseries 195} (1981) 467}.

\bibitem{Linde:1981mu}
A.~D. Linde, \emph{{A New Inflationary Universe Scenario: A Possible Solution
  of the Horizon, Flatness, Homogeneity, Isotropy and Primordial Monopole
  Problems}}, \href{https://doi.org/10.1016/0370-2693(82)91219-9}{\emph{Phys.
  Lett. B} {\bfseries 108} (1982) 389}.

\bibitem{Albrecht:1982wi}
A.~Albrecht and P.~J. Steinhardt, \emph{{Cosmology for Grand Unified Theories
  with Radiatively Induced Symmetry Breaking}},
  \href{https://doi.org/10.1103/PhysRevLett.48.1220}{\emph{Phys. Rev. Lett.}
  {\bfseries 48} (1982) 1220}.

\bibitem{Guth:1982ec}
A.~H. Guth and S.~Y. Pi, \emph{{Fluctuations in the New Inflationary
  Universe}}, \href{https://doi.org/10.1103/PhysRevLett.49.1110}{\emph{Phys.
  Rev. Lett.} {\bfseries 49} (1982) 1110}.

\bibitem{Martin:2013tda}
J.~Martin, C.~Ringeval and V.~Vennin, \emph{{Encyclopædia Inflationaris}},
  \href{https://doi.org/10.1016/j.dark.2014.01.003}{\emph{Phys. Dark Univ.}
  {\bfseries 5-6} (2014) 75} [\href{https://arxiv.org/abs/1303.3787}{{\ttfamily
  1303.3787}}].

\bibitem{Suzuki:2018cuy}
A.~Suzuki et~al., \emph{{The LiteBIRD Satellite Mission - Sub-Kelvin
  Instrument}}, \href{https://doi.org/10.1007/s10909-018-1947-7}{\emph{J. Low.
  Temp. Phys.} {\bfseries 193} (2018) 1048}
  [\href{https://arxiv.org/abs/1801.06987}{{\ttfamily 1801.06987}}].

\bibitem{Li:2015taa}
T.~Li, Z.~Li and D.~V. Nanopoulos, \emph{{Helical Phase Inflation via
  Non-Geometric Flux Compactifications: from Natural to Starobinsky-like
  Inflation}}, \href{https://doi.org/10.1007/JHEP10(2015)138}{\emph{JHEP}
  {\bfseries 10} (2015) 138}
  [\href{https://arxiv.org/abs/1507.04687}{{\ttfamily 1507.04687}}].

\bibitem{Li:2014vpa}
T.~Li, Z.~Li and D.~V. Nanopoulos, \emph{{Helical Phase Inflation}},
  \href{https://doi.org/10.1103/PhysRevD.91.061303}{\emph{Phys. Rev. D}
  {\bfseries 91} (2015) 061303}
  [\href{https://arxiv.org/abs/1409.3267}{{\ttfamily 1409.3267}}].

\bibitem{Li:2014unh}
T.~Li, Z.~Li and D.~V. Nanopoulos, \emph{{Helical Phase Inflation and Monodromy
  in Supergravity Theory}},
  \href{https://doi.org/10.1155/2015/397410}{\emph{Adv. High Energy Phys.}
  {\bfseries 2015} (2015) 397410}
  [\href{https://arxiv.org/abs/1412.5093}{{\ttfamily 1412.5093}}].

\bibitem{Freese:1990rb}
K.~Freese, J.~A. Frieman and A.~V. Olinto, \emph{{Natural inflation with pseudo
  - Nambu-Goldstone bosons}},
  \href{https://doi.org/10.1103/PhysRevLett.65.3233}{\emph{Phys. Rev. Lett.}
  {\bfseries 65} (1990) 3233}.

\bibitem{Ooguri:2006in}
H.~Ooguri and C.~Vafa, \emph{{On the Geometry of the String Landscape and the
  Swampland}},
  \href{https://doi.org/10.1016/j.nuclphysb.2006.10.033}{\emph{Nucl. Phys. B}
  {\bfseries 766} (2007) 21}
  [\href{https://arxiv.org/abs/hep-th/0605264}{{\ttfamily hep-th/0605264}}].

\bibitem{Ooguri:2018wrx}
H.~Ooguri, E.~Palti, G.~Shiu and C.~Vafa, \emph{{Distance and de Sitter
  Conjectures on the Swampland}},
  \href{https://doi.org/10.1016/j.physletb.2018.11.018}{\emph{Phys. Lett. B}
  {\bfseries 788} (2019) 180}
  [\href{https://arxiv.org/abs/1810.05506}{{\ttfamily 1810.05506}}].

\bibitem{Palti:2019pca}
E.~Palti, \emph{{The Swampland: Introduction and Review}},
  \href{https://doi.org/10.1002/prop.201900037}{\emph{Fortsch. Phys.}
  {\bfseries 67} (2019) 1900037}
  [\href{https://arxiv.org/abs/1903.06239}{{\ttfamily 1903.06239}}].

\bibitem{Landete:2018kqf}
A.~Landete and G.~Shiu, \emph{{Mass Hierarchies and Dynamical Field Range}},
  \href{https://doi.org/10.1103/PhysRevD.98.066012}{\emph{Phys. Rev. D}
  {\bfseries 98} (2018) 066012}
  [\href{https://arxiv.org/abs/1806.01874}{{\ttfamily 1806.01874}}].

\bibitem{Achucarro:2018vey}
A.~Achúcarro and G.~A. Palma, \emph{{The string swampland constraints require
  multi-field inflation}},
  \href{https://doi.org/10.1088/1475-7516/2019/02/041}{\emph{JCAP} {\bfseries
  02} (2019) 041} [\href{https://arxiv.org/abs/1807.04390}{{\ttfamily
  1807.04390}}].

\bibitem{Jaman:2018ucm}
N.~Jaman and K.~Myrzakulov, \emph{{Braneworld inflation with an effective
  $\alpha$-attractor potential}},
  \href{https://doi.org/10.1103/PhysRevD.99.103523}{\emph{Phys. Rev. D}
  {\bfseries 99} (2019) 103523}
  [\href{https://arxiv.org/abs/1807.07443}{{\ttfamily 1807.07443}}].

\bibitem{Lin:2018kjm}
C.-M. Lin, K.-W. Ng and K.~Cheung, \emph{{Chaotic inflation on the brane and
  the Swampland Criteria}},
  \href{https://doi.org/10.1103/PhysRevD.100.023545}{\emph{Phys. Rev. D}
  {\bfseries 100} (2019) 023545}
  [\href{https://arxiv.org/abs/1810.01644}{{\ttfamily 1810.01644}}].

\bibitem{Brahma:2018hrd}
S.~Brahma and M.~Wali~Hossain, \emph{{Avoiding the string swampland in
  single-field inflation: Excited initial states}},
  \href{https://doi.org/10.1007/JHEP03(2019)006}{\emph{JHEP} {\bfseries 03}
  (2019) 006} [\href{https://arxiv.org/abs/1809.01277}{{\ttfamily
  1809.01277}}].

\bibitem{Safsafi:2018cua}
A.~Safsafi, I.~Khay, F.~Salamate, H.~Chakir and M.~Bennai, \emph{{On Chaplygin
  Gas Braneworld Inflation with Monomial Potential}},
  \href{https://doi.org/10.1155/2018/2958605}{\emph{Adv. High Energy Phys.}
  {\bfseries 2018} (2018) 2958605}
  [\href{https://arxiv.org/abs/1804.11198}{{\ttfamily 1804.11198}}].

\bibitem{Es-sobbahi:2018yfh}
H.~Es-sobbahi and M.~Nach, \emph{{On braneworld inverse power-law inflation}},
  \href{https://doi.org/10.1142/S0217751X18500586}{\emph{Int. J. Mod. Phys. A}
  {\bfseries 33} (2018) 1850058}.

\bibitem{Bhattacharya:2019ryo}
S.~Bhattacharya, K.~Das and M.~R. Gangopadhyay, \emph{{Probing the era of
  reheating for reconstructed inflationary potential in the RS II braneworld}},
   \href{https://arxiv.org/abs/1908.02542}{{\ttfamily 1908.02542}}.

\bibitem{Jawad:2019hzo}
A.~Jawad, I.~Zehra and W.~Nazeer, \emph{{Warm vector inflation in brane-world
  scenario}}, \href{https://doi.org/10.1007/s10509-019-3518-z}{\emph{Astrophys.
  Space Sci.} {\bfseries 364} (2019) 30}.

\bibitem{Sabir:2019wel}
M.~Sabir, W.~Ahmed, Y.~Gong and Y.~Lu, \emph{{$\alpha$-attractor from
  superconformal E-models in brane inflation}},
  \href{https://doi.org/10.1140/epjc/s10052-019-7589-3}{\emph{Eur. Phys. J. C}
  {\bfseries 80} (2020) 15} [\href{https://arxiv.org/abs/1903.08435}{{\ttfamily
  1903.08435}}].

\bibitem{Sabir:2019bsh}
M.~Sabir, W.~Ahmed, Y.~Gong, S.~Hu, T.~Li and L.~Wu, \emph{{A note on brane
  inflation under consistency conditions}},
  \href{https://arxiv.org/abs/1905.03033}{{\ttfamily 1905.03033}}.

\bibitem{Alexander:2000xv}
S.~Alexander, R.~H. Brandenberger and D.~A. Easson, \emph{{Brane gases in the
  early universe}},
  \href{https://doi.org/10.1103/PhysRevD.62.103509}{\emph{Phys. Rev. D}
  {\bfseries 62} (2000) 103509}
  [\href{https://arxiv.org/abs/hep-th/0005212}{{\ttfamily hep-th/0005212}}].

\bibitem{Akrami:2018odb}
{\scshape Planck} collaboration, \emph{{Planck 2018 results. X. Constraints on
  inflation}},  \href{https://arxiv.org/abs/1807.06211}{{\ttfamily
  1807.06211}}.

\bibitem{Ade:2018gkx}
{\scshape BICEP2, Keck Array} collaboration, \emph{{BICEP2 / Keck Array x:
  Constraints on Primordial Gravitational Waves using Planck, WMAP, and New
  BICEP2/Keck Observations through the 2015 Season}},
  \href{https://doi.org/10.1103/PhysRevLett.121.221301}{\emph{Phys. Rev. Lett.}
  {\bfseries 121} (2018) 221301}
  [\href{https://arxiv.org/abs/1810.05216}{{\ttfamily 1810.05216}}].

\bibitem{Langlois:2000ns}
D.~Langlois, R.~Maartens and D.~Wands, \emph{{Gravitational waves from
  inflation on the brane}},
  \href{https://doi.org/10.1016/S0370-2693(00)00957-6}{\emph{Phys. Lett. B}
  {\bfseries 489} (2000) 259}
  [\href{https://arxiv.org/abs/hep-th/0006007}{{\ttfamily hep-th/0006007}}].

\bibitem{Kallosh:2015lwa}
R.~Kallosh and A.~Linde, \emph{{Planck, LHC, and $\alpha$-attractors}},
  \href{https://doi.org/10.1103/PhysRevD.91.083528}{\emph{Phys. Rev. D}
  {\bfseries 91} (2015) 083528}
  [\href{https://arxiv.org/abs/1502.07733}{{\ttfamily 1502.07733}}].

\bibitem{Kallosh:2013yoa}
R.~Kallosh, A.~Linde and D.~Roest, \emph{{Superconformal Inflationary
  $\alpha$-Attractors}},
  \href{https://doi.org/10.1007/JHEP11(2013)198}{\emph{JHEP} {\bfseries 11}
  (2013) 198} [\href{https://arxiv.org/abs/1311.0472}{{\ttfamily 1311.0472}}].

\bibitem{Shiromizu:1999wj}
T.~Shiromizu, K.-i. Maeda and M.~Sasaki, \emph{{The Einstein equation on the
  3-brane world}},
  \href{https://doi.org/10.1103/PhysRevD.62.024012}{\emph{Phys. Rev. D}
  {\bfseries 62} (2000) 024012}
  [\href{https://arxiv.org/abs/gr-qc/9910076}{{\ttfamily gr-qc/9910076}}].

\bibitem{Csaki:1999jh}
C.~Csaki, M.~Graesser, C.~F. Kolda and J.~Terning, \emph{{Cosmology of one
  extra dimension with localized gravity}},
  \href{https://doi.org/10.1016/S0370-2693(99)00896-5}{\emph{Phys. Lett. B}
  {\bfseries 462} (1999) 34}
  [\href{https://arxiv.org/abs/hep-ph/9906513}{{\ttfamily hep-ph/9906513}}].

\bibitem{Maartens:1999hf}
R.~Maartens, D.~Wands, B.~A. Bassett and I.~Heard, \emph{{Chaotic inflation on
  the brane}}, \href{https://doi.org/10.1103/PhysRevD.62.041301}{\emph{Phys.
  Rev. D} {\bfseries 62} (2000) 041301}
  [\href{https://arxiv.org/abs/hep-ph/9912464}{{\ttfamily hep-ph/9912464}}].

\bibitem{Binetruy:1999hy}
P.~Binetruy, C.~Deffayet, U.~Ellwanger and D.~Langlois, \emph{{Brane
  cosmological evolution in a bulk with cosmological constant}},
  \href{https://doi.org/10.1016/S0370-2693(00)00204-5}{\emph{Phys. Lett. B}
  {\bfseries 477} (2000) 285}
  [\href{https://arxiv.org/abs/hep-th/9910219}{{\ttfamily hep-th/9910219}}].

\bibitem{Binetruy:1999ut}
P.~Binetruy, C.~Deffayet and D.~Langlois, \emph{{Nonconventional cosmology from
  a brane universe}},
  \href{https://doi.org/10.1016/S0550-3213(99)00696-3}{\emph{Nucl. Phys. B}
  {\bfseries 565} (2000) 269}
  [\href{https://arxiv.org/abs/hep-th/9905012}{{\ttfamily hep-th/9905012}}].

\bibitem{Randall:1999vf}
L.~Randall and R.~Sundrum, \emph{{An Alternative to compactification}},
  \href{https://doi.org/10.1103/PhysRevLett.83.4690}{\emph{Phys. Rev. Lett.}
  {\bfseries 83} (1999) 4690}
  [\href{https://arxiv.org/abs/hep-th/9906064}{{\ttfamily hep-th/9906064}}].

\bibitem{Bento:2008yx}
M.~C. Bento, R.~G. Felipe and N.~M.~C. Santos, \emph{{Brane assisted
  quintessential inflation with transient acceleration}},
  \href{https://doi.org/10.1103/PhysRevD.77.123512}{\emph{Phys. Rev. D}
  {\bfseries 77} (2008) 123512}
  [\href{https://arxiv.org/abs/0801.3450}{{\ttfamily 0801.3450}}].

\bibitem{Okada:2015vka}
N.~Okada and Q.~Shafi, \emph{{$\mu$-term hybrid inflation and split
  supersymmetry}},
  \href{https://doi.org/10.1016/j.physletb.2017.11.015}{\emph{Phys. Lett. B}
  {\bfseries 775} (2017) 348}
  [\href{https://arxiv.org/abs/1506.01410}{{\ttfamily 1506.01410}}].

\bibitem{Kolb:1990vq}
E.~W. Kolb and M.~S. Turner, \emph{{The Early Universe}}, {\emph{Front. Phys.}
  {\bfseries 69} (1990) 1}.

\bibitem{Ahmed:2018jlv}
W.~Ahmed and A.~Karozas, \emph{{Inflation from a no-scale supersymmetric
  $SU(4)_{c}\times{SU(2)_{L}\times{SU(2)_{R}}}$ model}},
  \href{https://doi.org/10.1103/PhysRevD.98.023538}{\emph{Phys. Rev. D}
  {\bfseries 98} (2018) 023538}
  [\href{https://arxiv.org/abs/1804.04822}{{\ttfamily 1804.04822}}].

\end{thebibliography}

\providecommand{\href}[2]{#2}\begingroup\raggedright\endgroup

\end{document}